# Real-time interpretation of neutron vibrational spectra with symmetry-equivariant Hessian matrix prediction

Bowen Han,[1]* Pei Zhang,[2] Kshitij Mehta,[3] Massimiliano Lupo Pasini,[2] Mingda Li,[4] Yongqiang Cheng[1]*

[1]Neutron Scattering Division, Oak Ridge National Laboratory, Oak Ridge, TN, USA

[2]Computational Sciences and Engineering Division, Oak Ridge National Laboratory, Oak Ridge, TN, USA

[3]Computer Science and Mathematics Division, Oak Ridge National Laboratory, Oak Ridge, TN, USA

[4]Quantum Measurement Group, Massachusetts Institute of Technology, Cambridge, MA, USA

Emails: hanb1@ornl.gov, chengy@ornl.gov

**Abstract**

The vibrational behavior of molecules serves as a crucial fingerprint of their structure, chemical state, and surrounding environment. Neutron vibrational spectroscopy provides comprehensive measurements of vibrational modes without selection rule restrictions. However, analyzing and interpreting the resulting spectra remains a computationally formidable task. Here, we introduce a symmetry-aware neural network that directly predicts Hessian matrices from molecular structures, thereby enabling rapid vibrational spectral reconstruction. Unlike traditional approaches that focus on eigenvalue prediction, the Hessian matrix provides richer, more fundamental information with broader applications and superior extrapolation. This approach also paves the way for predicting other properties, such as reaction pathways. Trained on small molecules, our model achieves spectroscopic-level accuracy, allowing real-time, unambiguous peak assignment. Moreover, it maintains high accuracy for larger molecules, demonstrating strong transferability. This adaptability unlocks new capabilities, including on-the-fly spectral interpretation for future autonomous laboratories, and offers insights into molecular design for targeted chemical pathways.

*This manuscript has been authored in part by UT-Battelle, LLC, under contract DE-AC05-00OR22725 with the US Department of Energy (DOE). The US government retains and the publisher, by accepting the article for publication, acknowledges that the US government retains a nonexclusive, paid-up, irrevocable, worldwide license to publish or reproduce the published form of this manuscript, or allow others to do so, for US government purposes. DOE will provide public access to these results of federally sponsored research in accordance with the DOE Public Access Plan (http://energy.gov/downloads/doe-public-access-plan).*

**Introduction**

Vibrational spectroscopy using inelastic neutron scattering (INS) is a uniquely powerful tool to study molecular systems.[1] Unlike optical-based methods such as infrared (IR) and Raman spectroscopy, INS uses neutrons to probe the vibrational dynamics. This neutron-based approach offers several clear advantages. First, since neutrons directly interact with atomic nuclei, there are no selection rules as in IR/Raman, allowing INS to capture the full vibrational modes. Second, the energies of thermal neutrons are in the same range as typical molecular vibrations, making INS ideally suited to observe with high resolution a broad range of vibrational frequencies – including ultralow frequencies. Third, neutron scattering cross-section of H is almost one order of magnitude higher than the other common elements such as C, O, and N in molecules, making INS very sensitive to hydrogen-associated vibrational modes, with deuteration enhancing the selectivity. Fourth, the weak neutron-matter interaction allows a deeper penetration into materials for bulk sensitivity. Finally, the simulation of INS spectra is more straightforward compared to IR or Raman spectroscopy, enabling direct and quantitative comparisons between experiments and models, leading to more accurate conclusions and detailed insights.

The advantage of INS in measuring the low-to-intermediate frequency range in vibrational spectra, where peaks are often abundant and close to each other, also poses challenges in peak assignment and spectral interpretation. Therefore, accurately simulating the expected INS spectrum from a molecule is a critical requirement in data analysis and interpretation. The normal modes and the corresponding frequencies of a molecule can be solved by calculating the interatomic interactions and Hessian matrices. To obtain reliable vibrational properties, density functional theory (DFT) or more accurate methods are usually required. These methods are computationally expensive, especially for relatively large molecules, and they often become the bottleneck that slows down the data pipeline. Figure 1 illustrates the complementary role of INS, IR, and Raman, as well as the advantages of INS in revealing the low-to-intermediate frequency vibrational modes.[2] A simulated spectrum based on DFT demonstrates how rigorous peak matching and assignment to specific vibrational modes can be achieved when the simulation aligns reasonably well with the experiment. Unfortunately, to achieve sufficiently good agreement for unambiguous assignment, one usually needs high-accuracy DFT, which can take hours or longer for the structural optimization and vibration calculations.

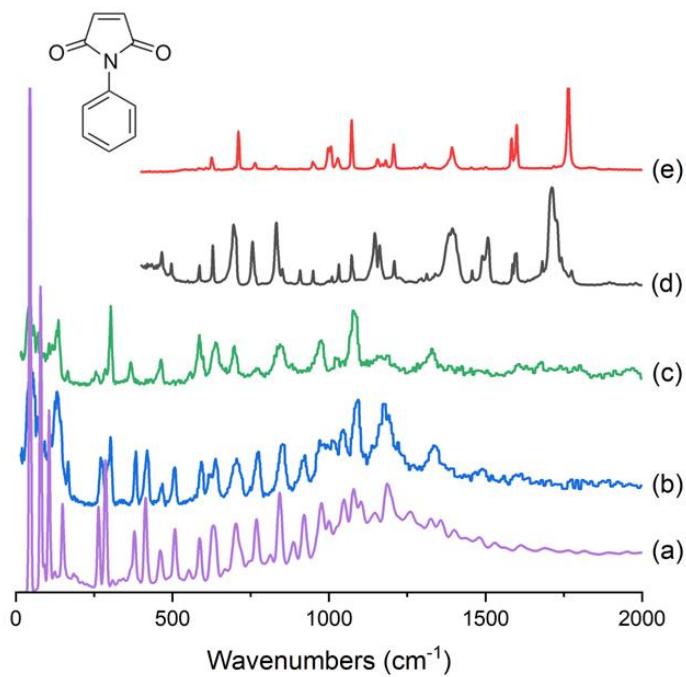

Figure 1. Vibrational spectra of N-phenylmaleimide. (a) Simulated INS spectrum based on DFT. (b) Experimentally measured INS spectrum. (c) Experimentally measured spectrum on partially deuterated N-phenylmaleimide (with the phenyl ring deuterated). (d) Experimentally measured IR spectrum. (e) Experimentally measured Raman spectrum. Corresponding peaks in (a) and (b) match each other so that the observed peaks can be assigned to the responsible vibrational modes solved in simulation. All experimental spectra were provided by Parker.[2]

Recent development in advanced data processing algorithms, especially using machine learning (ML), has opened the door to accelerating the analysis of INS data. Instead of simulating the INS spectra from scratch using DFT, direct prediction of INS spectra from the atomic structure has been demonstrated in periodic crystals.[3] However, similar prediction for molecular systems is proven more challenging, because the molecular INS spectra are usually rich in fine structures, with many sharp peaks in a narrow frequency range. This is in contrast to the smoother and broader phonon bands in periodic crystals. Therefore, instead of predicting the spectra directly, an alternative approach is to predict the Hessian matrices, from which the normal modes and frequencies can be calculated quickly, followed by the simulation of the INS spectra. Predicting the Hessian matrix offers significant advantages over directly predicting eigenvalues of the vibrational spectra or relying on ML interatomic potentials (MLIPs). This is so since the Hessian matrix encodes the full second-order energy derivatives with respect to atomic positions, capturing both vibrational frequencies (eigenvalues) and normal modes (eigenvectors). Unlike eigenvalue prediction, which is task-specific and loses intermediate structural details, the Hessian matrix provides a comprehensive foundation for deriving multiple properties, such as transitional states away from equilibrium. Compared to MLIPs, which focus on total energies and forces, Hessian prediction is more end-to-end and directly incorporates second-order information, which is expected to be much faster than any MLIP. Additionally, the symmetry constraints inherent to the Hessian ensure physically consistent predictions that generalize better across diverse materials and structures. By preserving this rich, interpretable information, Hessian prediction stands as a more versatile and robust approach for understanding material behavior. This approach also comes with additional flexibility to calculate other related properties, such as INS spectra measured on different instruments, with different energy range and resolution, other type of vibrational spectra (e.g., IR and Raman), as well as thermodynamic properties. Predicting the Hessian matrix from the molecular structure has been explored recently by several groups. Domenichini and Dellago presented a ML model based on a random forest to directly predict the molecular Hessian matrix in redundant internal coordinates.[4] The model was trained on a subset of QM7 dataset (6810 molecules) and tested on QM9 dataset. The efficient approach accurately predicted localized vibrations but struggled with non-localized and low-frequency modes. Zou et al. developed DetaNet, a deep-learning framework combining E(3)-equivariance and self-attention mechanism, to predict the molecular Hessian matrices for selected organic molecules.[5] They decomposed the Hessian matrix into an atomic tensor and an interatomic tensor and predicted them with separated models. Trained with 117,000 organic molecules in QM9S dataset, DetaNet achieved high accuracy ($R^2 = 0.9994$) within the QM9S dataset compared to the DFT calculation. Fang et al. used E(3)-equivariant graph neural networks (GNNs) to predict the molecular Hessian matrix based on the MLIP model.[6] In addition to the energy and force data, the molecular Hessian data are used in the training process, effectively improving the accuracy of force and the vibrational properties.

In this study, we use NequIP,[7] a symmetry-aware neural network based on e3nn,[8] to represent the molecular structure and predict the potential energy. We then solve the Hessian matrices by taking the second derivative of the potential energy with respect to the atomic coordinates using JAX. The DFT calculated Hessian matrices are used as ground truth to perform supervised training, such that the trained model can be further used as a DFT surrogate for the calculation of Hessian matrices. Compared to the existing ML based Hessian predictions, our method systematically exhibits high accuracy and transferability over a wide range of molecules. By focusing on the application and verification with INS, we highlight the capability of the model to predict complex spectra with high density of peaks in the low-to-intermediate energy range, as well as the implications for real-time spectral interpretation.

**Results**

Figure 2 shows a diagram illustrating the neural network architecture and the training and application workflow. We used the QM8 database[9, 10] containing over 20,000 small molecules (with no more than eight non-hydrogen atoms) to train the model. Details on the preparation of the QM8 database can be found in Methods. Our target is the INS spectra measured at the indirect geometry neutron spectrometers, such as TOSCA at ISIS[11] and VISION at SNS.[12] These spectrometers are optimized to measure molecular systems. INS spectra on other spectrometers or under other measurement conditions can also be obtained easily once the Hessian matrix is known.

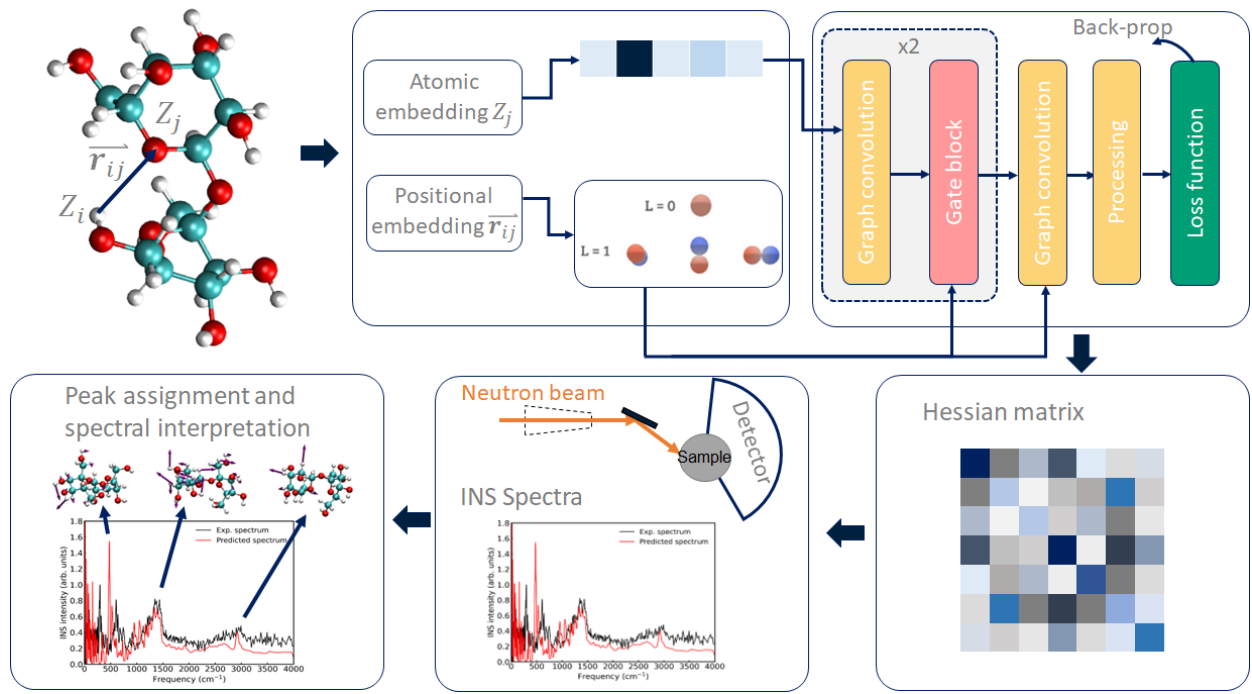

Figure 2. Architecture of the neural network and workflow for rapid calculation and interpretation of INS spectra from predicted Hessian matrices, with the molecular structure as the only input.

Figure 3 shows the excellent performance of the model on QM8 data, achieving the highest $R^2$ for Hessian matrix prediction. Specifically, the $R^2$ is as high as 0.9996 and almost the same for the validation and testing datasets, compared to the training datasets. This indicates that there is no significant overfitting, and the model is truly learning the features in the molecular structure that contribute to the Hessian matrix elements.

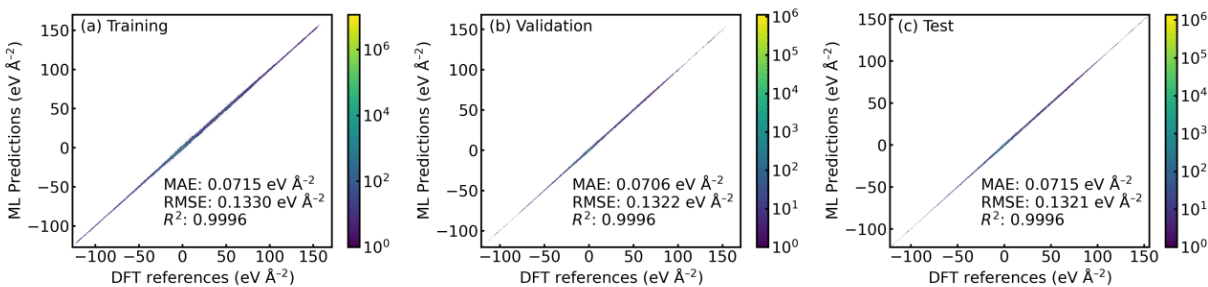

Figure 3. Correlation plots comparing predicted elements in molecular Hessian matrices with DFT reference values for (a) training, (b) validation, and (c) test data, all from the QM8 database. Color scale indicates the number of data points. Performance metrics, including MAE, RMSE, and $R^2$, demonstrate high accuracy across all datasets.

Using the predicted Hessian, we can readily calculate normal modes and the expected INS spectra. Figure 4 compares the spectra obtained from the DFT Hessian and the predicted Hessian. The dense and sharp peaks below 2000 cm$^{-1}$ can be clearly seen, and assignment of these peaks without accurate modeling is not possible. The comparison in Figure 4, with insets showing the details in the lower frequency range, confirms the prediction accuracy is sufficient to unambiguously match most predicted peaks to their DFT ground truth. Once the peaks are assigned, the responsible vibrational mode leading to the peak can be identified and visualized, from which feature interpretations and additional insight can be obtained.

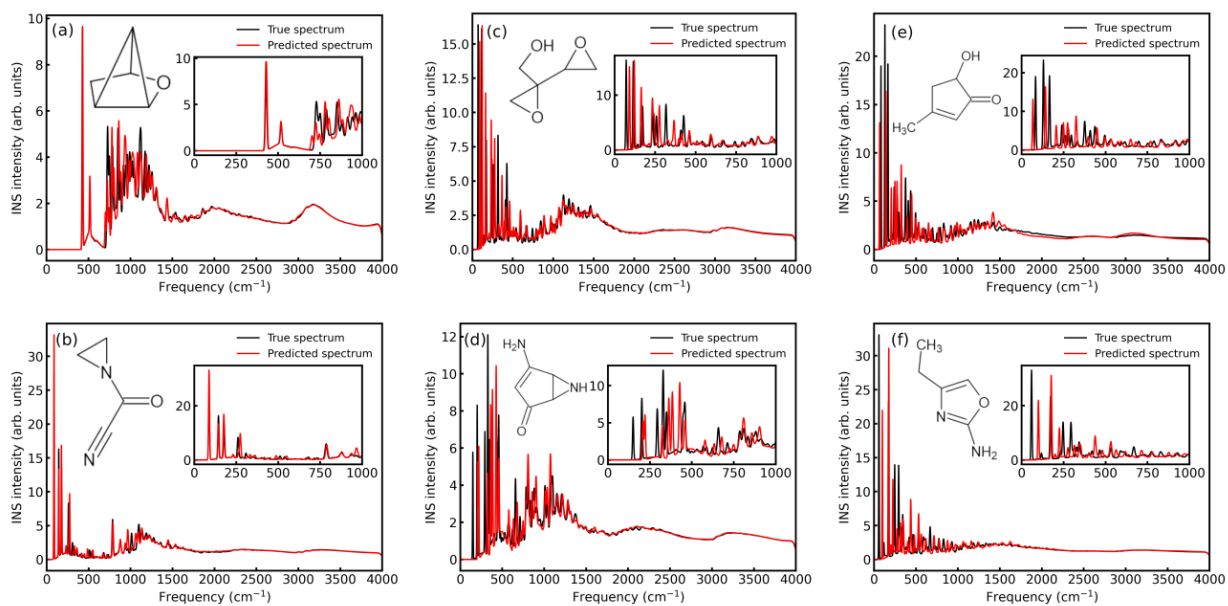

Figure 4. INS spectra of selected molecules in the QM8 database. Black lines represent DFT calculated (ground truth) spectra, and red lines represent spectra calculated using predicted Hessian matrices. Insets show details in the low frequency range. The corresponding molecular structure is also displayed. (a) and (b) are two examples among the molecules showing the best agreement (top 25%); (c) and (d) are two examples representing the average agreement (between 25% and 75%); (e) and (f) are two examples from the molecules showing worse agreement (bottom 25%). The criteria used to quantify agreement are explained later in the paper.

For small molecules, although DFT calculation is much more expensive than the neural network prediction, it can still be completed in a reasonable time. The benefits of the data-driven approach become more appreciable when applied to larger molecules. For that, we will have to demonstrate the transferability of the model. In Figure 5, we first use our model trained on QM8 molecules to study the slightly larger molecules in the QM9 database.[13] The accuracy in Hessian prediction, as indicated by the MAE, RMSE, and $R^2$, is essentially on the same level (even slightly better, with an $R^2$ of 0.9998). We then scanned the PubChem database[14] for some significantly larger molecules. These molecules contain up to 30 non-hydrogen atoms (200 atoms in total) and require many hours to calculate using DFT (at the same accuracy level as we used for QM8). We verified our model prediction with the DFT results; the performance is still excellent with only a minimum decrease from QM8/QM9 ($R^2$=0.998). Note that the model was trained only with the QM8 dataset, and the QM9 or PubChem molecules are never used in the training.

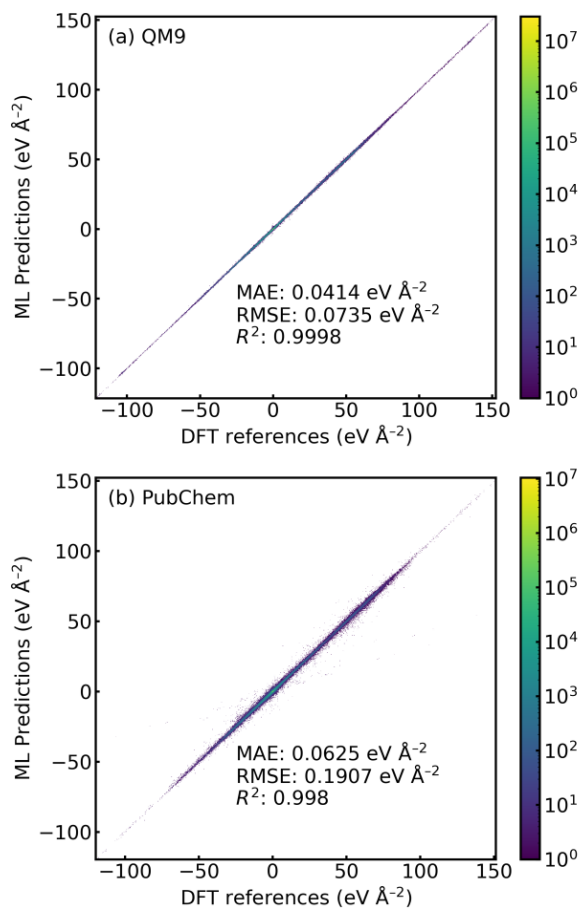

Figure 5. Correlation plots comparing predicted elements in molecular Hessian matrices with DFT reference values for (a) selected molecules in QM9 database and (b) selected molecules in PubChem database. Performance metrics, including MAE, RMSE, and $R^2$, demonstrate high accuracy, even though all molecules are larger than those used in the model training (from QM8 database).

The exceptional performance in Hessian prediction is expected to translate to excellent agreement in the INS spectra, as indeed shown in Figure 6. Despite the reduced spacing between peaks in the low-to-intermediate frequency range in larger molecules, it is still possible to confidently assign most of the peaks.

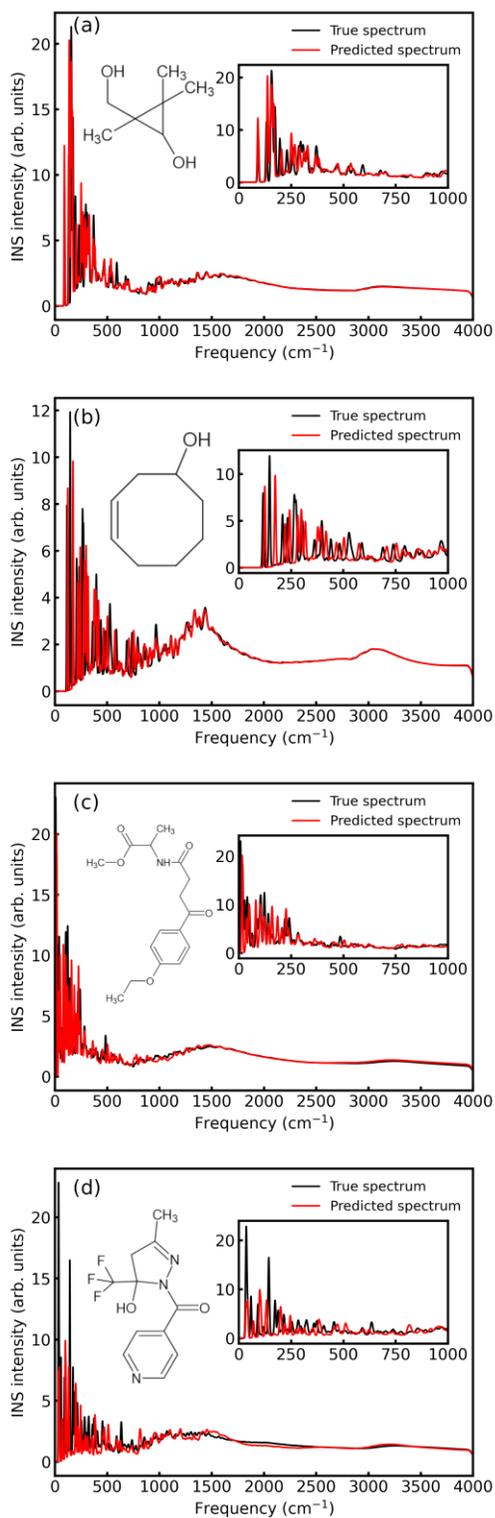

Figure 6. INS spectra of selected molecules in the QM9, (a) and (b), and PubChem, (c) and (d), databases. Black lines represent DFT calculated (ground truth) spectra, and red lines represent spectra calculated using predicted Hessian matrices. Insets show details in the low frequency range. The corresponding molecular structure is also displayed.

In general, peaks in the simulated and experimental INS spectra do not need to align perfectly for assignment. The peak position, shape, and intensity can be off, but the overall profile often tells us which peaks are of the same origin. To further quantify the "interpretability" or "assignability" of the spectra and provide some statistical information on the entire dataset, we calculated the Spearman coefficient[15] and plot its distribution in QM8, QM9, and PubChem molecules, as shown in Figure 7. The Spearman coefficient describes how well two datasets are "synchronized" with each other. We have inspected spectra pairs with Spearman coefficient in a wide range, and found that, as a rule of thumb, if the Spearman coefficient is larger than 0.6, the two spectra are generally aligned, and most of the peaks are assignable. Figure 7 illustrates that when only fundamental excitations are included in the spectra, they tend to be very well aligned, showing that the predictions of normal modes and frequencies are highly accurate. Including higher order excitations (to account for multi-phonon processes) and wing calculations (to mimic intermolecular modes in solid state)[16] can worsen the alignment, due to error propagation and the wing shape approximation. However, the agreement is still sufficiently good in most cases, as demonstrated in Figure 4 and Figure 6. Specifically, Figures 4(e) and 4(f) are two cases with Spearman coefficient of ~0.6, which are considered marginally assignable. Over 95% of cases in all datasets have Spearman coefficient higher than 0.6.

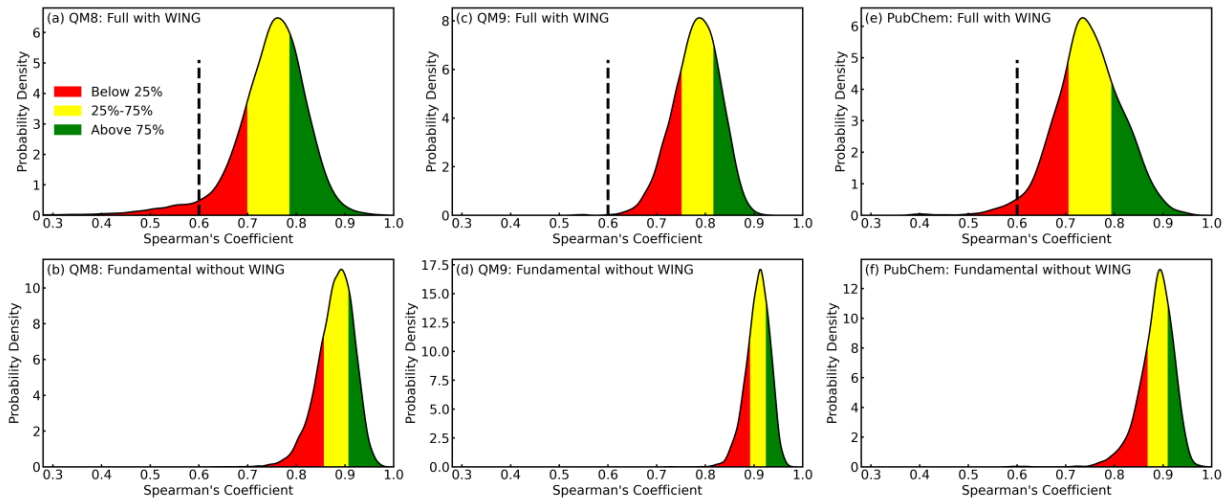

Figure 7. Distribution of Spearman coefficients in QM8, QM9, and PubChem datasets. Higher Spearman coefficient means better alignment between the predicted and ground truth spectra. The example spectra in Figure 4 were chosen from the top 25% (green), 25-75% (yellow), and bottom 25% (red), respectively. Dashed line marks a rough cutoff above which the two spectra are generally aligned, and peaks are assignable.

Based on the above analysis, we expect the model can help to achieve real-time peak assignment and data interpretation in molecular spectroscopy. Of course, when the model is used in an application scenario, there will be no DFT results to compare with. Instead, the model prediction will be directly compared with the experimental spectra, and the observed peaks will be assigned to the corresponding simulated ones, from which the origin of the peaks can be revealed, and the responsible vibrational modes can be visualized. To illustrate how it works, in Figure 8 we show two examples of triptindane and sucrose. The overall agreement between prediction and experiment is excellent, considering that the two molecules are relatively large, and the prediction was made rapidly with the molecular structure as the only input. Specifically, almost all peaks in the case of triptindane are aligned and can be easily assigned. With sucrose, the agreement at above 1000 cm$^{-1}$ is great, but less satisfactory at below 1000 cm$^{-1}$. Part of the reason is that the sample in the experiment is a solid, while the prediction was based on training data generated from single molecules. This discrepancy has a more significant impact on the OH groups in the sucrose molecule. Even so, assignment can still be performed if we keep these differences in mind. For example, the 600 cm$^{-1}$ peak seen in the experimental spectrum in sucrose is likely the sharp peak at 500 cm$^{-1}$ seen in the predicted one, corresponding to the OH torsion mode. The intermolecular interaction in the solid hardened this mode and made it broader.

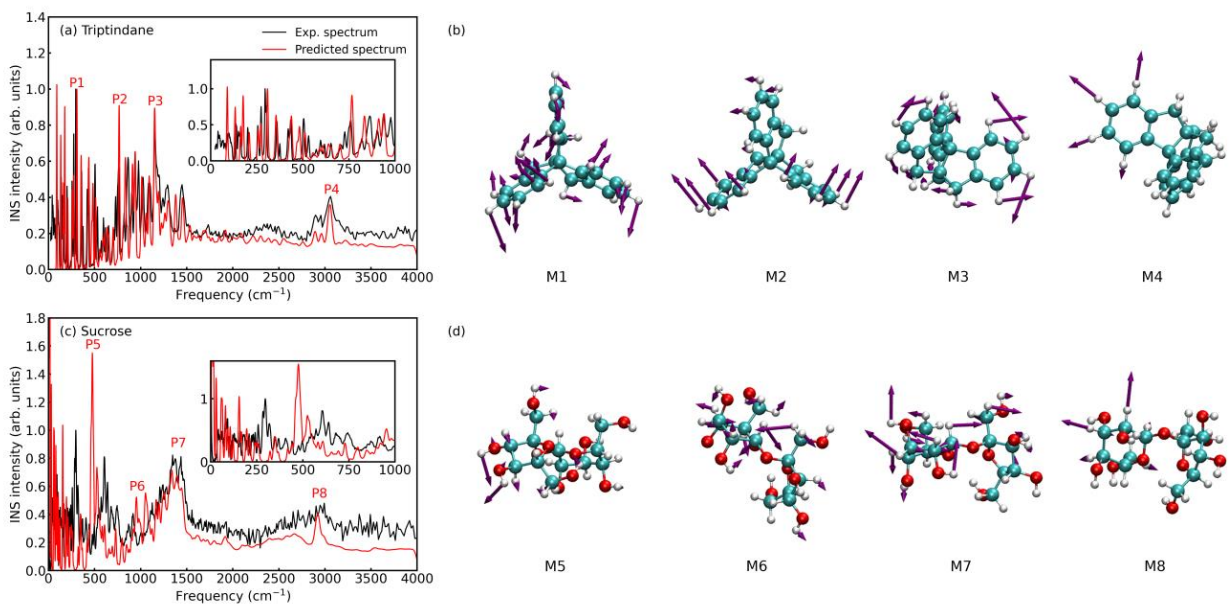

Figure 8. Comparison with experimental spectra for triptindane (measured at TOSCA)[17, 18] and sucrose (measured at VISION). Examples of peak assignment and model visualization are also shown. P1 to P8 in (a) and (c) correspond to M1 and M8, respectively.

**Discussion**

The entire workflow established above will only take seconds for molecules of small to medium sizes, in contrast to hours or even longer with the conventional DFT approach, with comparable accuracy in the resulting Hessian matrices and INS spectra. Another comparison we can make is with the molecular mechanics using empirical force fields. In this approach, interatomic interactions are calculated quickly from either analytical equations or tabulated forms. The speed can be high, but the accuracy is usually not sufficient for spectroscopy analysis. We show in Figure 9 the Spearman's coefficients and sample INS spectra calculated using the general AMBER force field (GAFF).[19] Figure 9(a) shows that the average Spearman's coefficient is considerably lower compared to the neural network model, and about 1/3 of the molecules have their Spearman's coefficient smaller than 0.6, meaning peak assignment will be challenging. Figure 9(b) shows a specific case where the errors are significant compared to DFT (and the ML model), and several major peaks cannot be assigned.

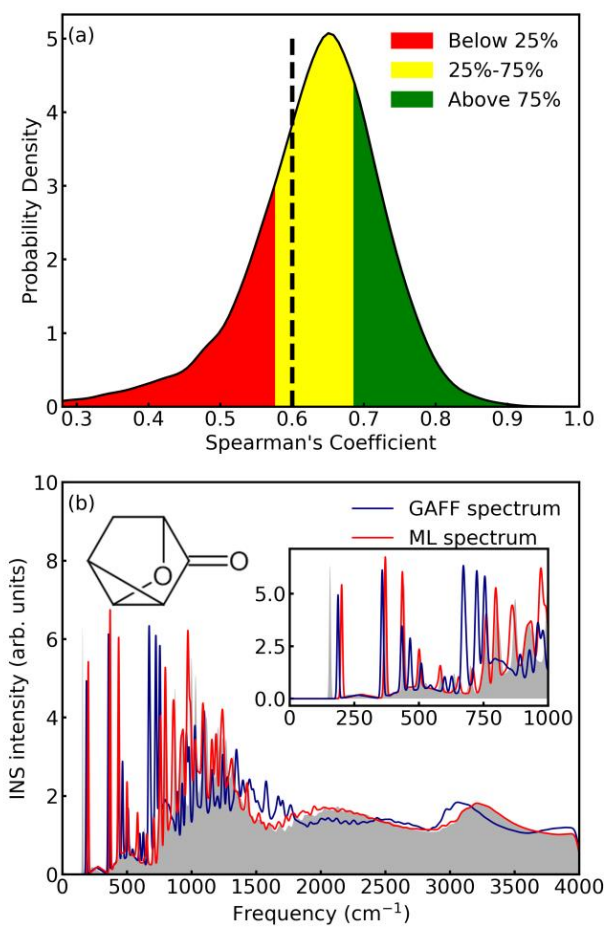

Figure 9. (a) Distribution of Spearman coefficients for spectra calculated using GAFF. (b) Selected INS spectrum calculated using GAFF, in comparison with our ML model and DFT ground truth (gray area).

Although the model trained with QM8 seems to be quite accurate, it can still be improved by including larger molecules in the training dataset. It is also possible to fine-tune the model parameters using experimental spectra, although one should be careful about the consistency issues when mixing DFT and experimental data. Alternatively, one could also take the predicted Hessian as the starting point and optimize the Hessian elements using experimental spectra. The optimized Hessian reproducing the experimental data will provide useful information on the interatomic interactions in the molecules.

Finally, while the model is trained with small molecules and applied to larger ones, there is a limit on its applicability. For super large molecules, such as proteins or biomolecules, different parts of the molecule may start to interact with each other through van der Waals interactions. This contribution is not included in the QM8 database and therefore will not be accounted for in the model. Larger errors are thus expected when applying the model to very large molecules.

In summary, we have demonstrated rapid and accurate prediction of INS spectra of molecular structures by using Hessian matrices, which, in turn, are predicted by symmetry-aware neural networks. Comparisons with experimentally measured spectra enable real-time peak assignment, mode visualization, and result interpretation. This capability is pivotal for steering experiments as data are being collected. When combined with active learning and other AI techniques, it brings us closer to achieving fully autonomous vibrational spectroscopy. Going beyond vibrational spectroscopy, the prediction of molecular Hessians has far-reaching implications. Accurate Hessian matrices enable the modeling of reaction pathways, critical for understanding chemical reactions in fields like catalysis, drug design, and material degradation. In energy applications, such as fuel cells and batteries, Hessian predictions can optimize molecular-level interactions and vibrational dynamics for enhanced performance and stability. Furthermore, they can facilitate the discovery of novel compounds in drug development and guide the design of corrosion-resistant materials by predicting key vibrational modes involved in chemical reactivity.

**Methods**

Database generation:

The training dataset of molecular Hessian matrices was obtained based on the QM8 database,[9, 10] which contains small molecules with no more than 8 heavy atoms. Starting with the original molecular structures in the database, geometry optimization and vibrational analysis were performed via Gaussian,[20] with B3LYP/6-311++G(d,p) level of theory. The Hessian database contains 21,693 molecules (93 out of the 21,786 molecules in the original QM8 database failed to converge in the geometry optimization. To validate the transferability of the model, we selected over 9000 molecules that have more than 23 atoms from QM9 database and 1644 randomly selected molecules with more than 9 non-hydrogen atoms from PubChem database. The DFT/ground-truth molecular Hessian matrices of these molecules were obtained through the same process used for QM8 database. The INS spectra of the molecules were simulated by OCLIMAX,[16] using the normal mode frequencies and eigenvectors solved from the molecular Hessian matrices.

Neural network and training:

We employed an E(3)-equivariant GNN[21], Neural Equivariant Interatomic Potentials (NequIP),[7] to predict the Hessian. We built the GNN with JAX and e3nn-jax.[8, 22] The layers in the GNN were implemented with nequip-jax.[23] QM8 database was randomly divided into training, validation and test datasets with a ratio of 90%, 5% and 5%. Early stopping technique was used to prevent overfitting (Figure 10). The model with lowest validation loss was saved for further production.

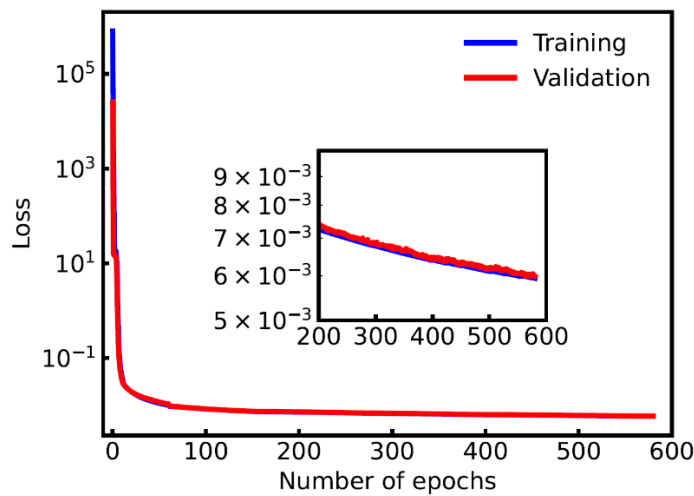

Figure 10. Evolution of the training and validation loss functions during the training of the model.

Spearman coefficient:

To compare the spectra obtained from DFT calculations/experiments to the ones generated from ML prediction, we use Spearman's rank correlation coefficient as the statistical measurement. Spearman's rank correlation coefficient is expressed as:

$$\rho = 1 - \frac{6\sum_i d_i^2}{n(n^2-1)} \quad ,$$

where $d_i$ is the difference between the ranks of $y_i$ and $\hat{y}_i$ in reference dataset and predicted dataset, respectively, and $n$ the number of elements in each vector. Spearman's rank correlation coefficients have been used to quantitatively compare the spectra of the same material but obtained from different sources.[5, 15, 24-26]

**Data availability**

The datasets used in this research are available at Zenodo (https://zenodo.org/records/14796533).

**Code availability**

The code is available at github (https://github.com/maplewen4/INS_molecule).

**Acknowledgements**

B.H. and Y.C. were supported by the Scientific User Facilities Division, Office of Basic Energy Sciences, U.S. Department of Energy (DOE), under Contract No. DE-AC0500OR22725 with UT Battelle, LLC. This research was partially sponsored by the Artificial Intelligence Initiative as part of the Laboratory Directed Research and Development (LDRD) program of Oak Ridge National Laboratory (ORNL). M.L. was partially supported by DOE, Office of Science, Basic Energy Sciences, Award No. DE-SC0021940. The computing resources were made available through the VirtuES and the ICE-MAN projects, funded by LDRD and Compute and Data Environment for Science (CADES) at ORNL. INS spectrum of sucrose was obtained at the VISION beamline of the Spallation Neutron Source, a DOE Office of Science User Facility operated by ORNL.

**References**

(1) Mitchell, P. C. H.; Parker, S. F.; Ramirez-Cuesta, T. A. J.; Tomkinson, J. *Vibrational Spectroscopy With Neutrons - With Applications In Chemistry, Biology, Materials Science And Catalysis*; World Scientific Publishing Company, 2005.
(2) Parker, S. F. Vibrational spectroscopy of N-phenylmaleimide. *Spectrochimica Acta Part A: Molecular and Biomolecular Spectroscopy* **2006**, *63* (3), 544-549. DOI: https://doi.org/10.1016/j.saa.2005.06.001.
(3) Cheng, Y.; Wu, G.; Pajerowski, D. M.; Stone, M. B.; Savici, A. T.; Li, M.; Ramirez-Cuesta, A. J. Direct prediction of inelastic neutron scattering spectra from the crystal structure. *Machine Learning: Science and Technology* **2023**, *4* (1), 015010. DOI: 10.1088/2632-2153/acb315.
(4) Domenichini, G.; Dellago, C. Molecular Hessian matrices from a machine learning random forest regression algorithm. *The Journal of Chemical Physics* **2023**, *159* (19), 194111. DOI: 10.1063/5.0169384.
(5) Zou, Z.; Zhang, Y.; Liang, L.; Wei, M.; Leng, J.; Jiang, J.; Luo, Y.; Hu, W. A deep learning model for predicting selected organic molecular spectra. *Nature Computational Science* **2023**, *3* (11), 957-964. DOI: 10.1038/s43588-023-00550-y.
(6) Fang, S.; Geiger, M.; Checkelsky, J. G.; Smidt, T. *Phonon predictions with E(3)-equivariant graph neural networks*; 2024. DOI: https://arxiv.org/abs/2403.11347.
(7) Batzner, S.; Musaelian, A.; Sun, L.; Geiger, M.; Mailoa, J. P.; Kornbluth, M.; Molinari, N.; Smidt, T. E.; Kozinsky, B. E(3)-equivariant graph neural networks for data-efficient and accurate interatomic potentials. *Nature Communications* **2022**, *13* (1), 2453. DOI: 10.1038/s41467-022-29939-5.
(8) Geiger, M.; Smidt, T. *e3nn: Euclidean Neural Networks*; 2022. DOI: https://arxiv.org/abs/2207.09453.

(9) Cheng, Y.; Stone, M. B.; Ramirez-Cuesta, A. J. A database of synthetic inelastic neutron scattering spectra from molecules and crystals. *Scientific Data* **2023**, *10* (1), 54. DOI: 10.1038/s41597-022-01926-x.
(10) Ruddigkeit, L.; van Deursen, R.; Blum, L. C.; Reymond, J.-L. Enumeration of 166 Billion Organic Small Molecules in the Chemical Universe Database GDB-17. *Journal of Chemical Information and Modeling* **2012**, *52* (11), 2864-2875. DOI: 10.1021/ci300415d.
(11) *TOSCA at ISIS*. https://www.isis.stfc.ac.uk/Pages/tosca.aspx (accessed 2024/11/22).
(12) *VISION at SNS*. https://neutrons.ornl.gov/vision (accessed 2024/11/22).
(13) Ramakrishnan, R.; Dral, P. O.; Rupp, M.; von Lilienfeld, O. A. Quantum chemistry structures and properties of 134 kilo molecules. *Scientific Data* **2014**, *1* (1), 140022. DOI: 10.1038/sdata.2014.22.
(14) Kim, S.; Chen, J.; Cheng, T.; Gindulyte, A.; He, J.; He, S.; Li, Q.; Shoemaker, B. A.; Thiessen, P. A.; Yu, B.; et al. PubChem 2023 update. *Nucleic Acids Research* **2023**, *51* (D1), D1373-D1380. DOI: 10.1093/nar/gkac956 (acccessed 11/22/2024).
(15) Henschel, H.; Andersson, A. T.; Jespers, W.; Mehdi Ghahremanpour, M.; van der Spoel, D. Theoretical Infrared Spectra: Quantitative Similarity Measures and Force Fields. *Journal of Chemical Theory and Computation* **2020**, *16* (5), 3307-3315. DOI: 10.1021/acs.jctc.0c00126.
(16) Cheng, Y. Q.; Daemen, L. L.; Kolesnikov, A. I.; Ramirez-Cuesta, A. J. Simulation of Inelastic Neutron Scattering Spectra Using OCLIMAX. *Journal of Chemical Theory and Computation* **2019**, *15* (3), 1974-1982. DOI: 10.1021/acs.jctc.8b01250.
(17) Parker, S. F.; Zhong, L.; Harig, M.; Kuck, D. Spectroscopic characterisation of centropolyindanes. *Physical Chemistry Chemical Physics* **2019**, *21* (8), 4568-4577, 10.1039/C8CP07311B. DOI: 10.1039/C8CP07311B.
(18) Parker, S. F. *TOSCA INS database*. https://www.isis.stfc.ac.uk/Pages/INS-database.aspx (accessed 2024/10/31).
(19) Wang, J.; Wolf, R. M.; Caldwell, J. W.; Kollman, P. A.; Case, D. A. Development and testing of a general amber force field. *Journal of Computational Chemistry* **2004**, *25* (9), 1157-1174. DOI: https://doi.org/10.1002/jcc.20035 (acccessed 2024/11/22).
(20) *Gaussian 16 Rev. C.01*; Wallingford, CT, 2016. (accessed 2024/11/22).
(21) Thomas, N.; Smidt, T.; Kearnes, S.; Yang, L.; Li, L.; Kohlhoff, K.; Riley, P. *Tensor field networks: Rotation- and translation-equivariant neural ne tworks for 3D point clouds*; 2018. DOI: https://arxiv.org/abs/1802.08219.
(22) *e3nn-jax*. https://github.com/e3nn/e3nn-jax (accessed 2024/11/22).
(23) *nequip-jax*. https://github.com/mariogeiger/nequip-jax (accessed 2024/11/22).
(24) Baumann, K.; Clerc, J. T. Computer-assisted IR spectra prediction — linked similarity searches for structures and spectra. *Analytica Chimica Acta* **1997**, *348* (1), 327-343. DOI: https://doi.org/10.1016/S0003-2670(97)00238-9.
(25) Ye, S.; Zhong, K.; Zhang, J.; Hu, W.; Hirst, J. D.; Zhang, G.; Mukamel, S.; Jiang, J. A Machine Learning Protocol for Predicting Protein Infrared Spectra. *Journal of the American Chemical Society* **2020**, *142* (45), 19071-19077. DOI: 10.1021/jacs.0c06530.
(26) Esch, B. v. d.; Peters, L. D. M.; Sauerland, L.; Ochsenfeld, C. Quantitative Comparison of Experimental and Computed IR-Spectra Extracted from Ab Initio Molecular Dynamics. *Journal of Chemical Theory and Computation* **2021**, *17* (2), 985-995. DOI: 10.1021/acs.jctc.0c01279.